# A STATISTICAL FRAMEWORK FOR TESTING FUNCTIONAL CATEGORIES IN MICROARRAY DATA

By William T. Barry, Andrew B. Nobel and Fred A. Wright

*Duke University Medical Center, University of North Carolina and University of North Carolina*

Ready access to emerging databases of gene annotation and functional pathways has shifted assessments of differential expression in DNA microarray studies from single genes to groups of genes with shared biological function. This paper takes a critical look at existing methods for assessing the differential expression of a group of genes (functional category), and provides some suggestions for improved performance. We begin by presenting a general framework, in which the set of genes in a functional category is compared to the complementary set of genes on the array. The framework includes tests for overrepresentation of a category within a list of significant genes, and methods that consider continuous measures of differential expression. Existing tests are divided into two classes. Class 1 tests assume gene-specific measures of differential expression are independent, despite overwhelming evidence of positive correlation. Analytic and simulated results are presented that demonstrate Class 1 tests are strongly anti-conservative in practice. Class 2 tests account for gene correlation, typically through array permutation that by construction has proper Type I error control for the induced null. However, both Class 1 and Class 2 tests use a null hypothesis that all genes have the same degree of differential expression. We introduce a more sensible and general (Class 3) null under which the profile of differential expression is the same within the category and complement. Under this broader null, Class 2 tests are shown to be conservative. We propose standard bootstrap methods for testing against the Class 3 null and demonstrate they provide valid Type I error control and more power than array permutation in simulated datasets and real microarray experiments.

**1. Introduction.** DNA microarrays allow researchers to simultaneously measure the coexpression of thousands of genes. They are widely used in biology and medicine to study the relationships between transcriptional expression and cellular processes or disease states. A primary application of microarrays is the identification of genes with differing expression across experimental conditions, or having significant association with a clinical outcome. Hereafter we will generically refer to the condition or clinical outcome as the *response* for each array, and the association between expression and response as differential expression (DE).

Analyses of DE often proceed in a gene-by-gene manner, in which the association between the response and the expression of each gene is assessed individually. A variety of methods have been proposed, including standard parametric tests, permutation and resampling-based methods, and Bayesian techniques [Dudoit et al. (2002), Newton et al. (2004) and Tusher, Tibshirani and Chu (2001)]. Using these methods, investigators can produce a ranked list of genes significantly associated to the response that may account for multiple testing through control of the family-wise error rate (FWER) or false discovery rate (FDR).

Although gene-specific analyses have yielded tremendous insight into the role of individual genes, they do not provide a mechanism for identifying larger-scale biological phenomena. With the ready availability of comprehensive annotation databases such as Gene Ontology (GO) [Ashburner et al. (2000)], researchers can now explore the coordinated involvement of *gene categories*, namely, sets of genes with shared annotation or function. A general framework is warranted for evaluating methods that test the associations of an entire category to the response of interest, and will allow a more systematic understanding of DE across the genome.

Beginning with Virtaneva et al. (2001), a number of procedures have been presented as ways to assess the association between a response and the expression of a gene category. The most commonly used tests begin with a list of genes deemed significant and look for over-representation of category members within the gene-list, using Fisher's Exact Test or other tests of association for $2 \times 2$ contingency tables [see Barry, Nobel and Wright (2005) for a list of references]. Other approaches more directly use the gene-specific measures of DE, rather than collapsing the data to the dichotomous outcome of significant association with the response. In these methods tests are constructed to compare the association of genes using the average differences of gene-specific statistics [Kim and Volsky (2005) and Boorsma et al. (2005)], or rank-based procedures for two-sample comparisons [Mootha et al. (2003), Barry, Nobel and Wright (2005) and Ben-shaul, Bergman and Soreq (2005)].

Gene category testing is now widely performed, and results are frequently reported without independent verification. As pointed out in a recent review by Allison, Cui, Page et al. (2006), even fundamental issues such as the formal definition of the underlying null hypothesis and a proper analysis of Type I error have not been provided for many of the methods in the literature. Recent work by Goeman and Buhlmann (2007) addresses some of the issues surrounding the assumptions of independence in gene category testing, while suggesting that commonly used methods fail to test for a more direct hypothesis of DE among the category members. Dudoit et al. (2007) have also described a very general framework for hypothesis testing that captures most existing methods for testing gene categories and proposes bootstrap-based testing. However, there remains a clear need to differentiate among existing methods in order to examine their strengths and potential deficiencies, and to place gene category testing on a firm statistical foundation.

1.1. *Contributions.* In this paper we provide a careful, extended examination of gene category testing and discuss how a standard application of bootstrap methodology offers improved performance and flexibility over some of the existing methods in the literature. We begin by defining a framework for gene category testing, which is general enough to include the majority of the existing methods in the literature. Within our framework, existing gene category methods can be divided into two distinct classes of procedures as defined by the following null hypotheses:

1. Gene-specific statistics are independent and identically distributed;
2. Gene-specific statistics follow a common null distribution, though they may be dependent.

Several shortcomings of these null hypotheses are demonstrated through analytic derivations and simulations using an example dataset.

We propose a broader null hypothesis that allows for arbitrary dependence between the expression of different genes, as well as varying degrees of association between the expression of a given gene and the response. Under this more general null, array permutation approaches can be quite conservative. The conservativeness can be explained in part through an analytical argument which shows that the maximum variance of the category-wide test statistic occurs under the special case induced by array permutation. To remedy this problem, we suggest a simple bootstrap-based test that is consistent with the general null hypothesis. We demonstrate the utility of the bootstrap test on a breast cancer dataset, and discuss other advantages that bootstrap-based tests have over array permutation procedures.

**2. Notation and general framework for gene category tests.** Let **x** be an $m \times n$ matrix containing the observed expression data for an experiment with $m$ genes and $n$ arrays. Let $x_{ij}$ be the element of the matrix corresponding to the $i$th gene in the $j$th array. The expression profile for gene $i$ is the row vector $\mathbf{x}_{i*}$, and the expression values of array $j$ are represented by the column vector $\mathbf{x}_{*j}$. We use lowercase letters to denote observed values, and uppercase (i.e., $\mathbf{X}$, $X_{ij}$, $\mathbf{X}_{i*}$, and $\mathbf{X}_{*j}$) to denote random versions of these quantities. The array-specific response information is denoted by $\mathbf{y}$, with element $y_j$ corresponding to array $j$. The response can be categorical (e.g., tumor grade or experimental group assignment) or continuous (e.g., survival time), and could potentially be multivariate. A category is represented by a subset $C \subseteq \{1,\ldots,m\}$ such that $i \in C$ if and only if gene $i$ is a member of the category. The size of a category $C$ will be denoted by $m_C = \sum_{i=1}^{m} I\{i \in C\}$. For any category $C$, its complement will be denoted by $\bar{C}$, and is of size $m_{\bar{C}} = m - m_C$.

We adopt the terminology of Barry, Nobel and Wright (2005), where it is noted that hypothesis tests of gene categories can be viewed as two-stage procedures (see Box 1). In the first stage, a *local statistic* measures the association between the expression profile of each gene and the response. We denote the local statistic of gene $i$ by $T_i = T(\mathbf{X}_{i*}, \mathbf{y})$ and let $t_i$ be the corresponding value based on observed data. In a two-condition experiment, the local statistic might be a $t$-statistic or average fold change. For more complex datasets, such as those with censored survival data, a local statistic derived from a Cox proportional hazard model may be used to test for association between gene expression and patient outcome. In many cases, $T$ is an estimate of an underlying gene-specific parameter that governs the association between response and expression. In the two-condition example above, the related parameters would be a scaled difference of means and a ratio of population means, respectively. Properties of local statistics are examined more fully in Section 5.3.

In the second stage of a gene category test, a *global statistic* examines the differential expression within the gene category through the collection of local statistics. The global statistic can be generally denoted by $U = U(T_1,\ldots,T_m:C)$, and in the following sections we describe many of the functional forms of $U(\cdot)$ that have been utilized in the literature. In the most commonly employed tests of gene categories, $U(\cdot)$ compares the local statistics of genes within a category $C$ to those in its complement. Methods focus on either detecting a difference in the proportion of genes with significant DE, or determining a shift in the average local statistic of the category against its complement. Goeman and Buhlmann (2007) have argued that comparing a category to its complement creates an unnecessary conflict between these methods and the gene-specific tests. However, alternatively proposed methods that directly test the DE within a category have their own drawbacks. For many direct tests, the null hypothesis will tend to be rejected more often for large categories than small categories. This will be true even if the genes in the category are chosen at random. For this reason, we will limit our focus to tests that compare a category to its complement.

Among the current methods for analyzing gene categories, there are various ways to classify the tests that have been proposed (e.g., by the choice of global statistic). In terms of Type I error control, we argue that the more meaningful distinction is based on the implicit null hypotheses, as described in the following sections. Most existing procedures can be roughly divided according to whether *array permutation* is used, but we note that additional requirements must be placed on the local statistics in order for the inference to be sensible.

Throughout our paper we treat the categories to which a gene belongs as a fixed property of the gene.

---

**Box 1:** Common elements of gene category tests

Gene category tests are typically two-stage procedures requiring the following statistics:
- A *local statistic* that measures the association between the response (e.g., experimental condition) and the expression of each gene.
- A *global statistic* that examines the local statistics within a category, often in comparison to those of its complement.

For each global statistic there are two broad classes of hypothesis tests:
1. Parametric or rank-based procedures that assume independent and identically distributed local statistics, or alternatively, gene permutation methods that induce the same approximate null.
2. Array permutation methods which maintain the correlation structure while inducing a null of no associations with the response.

Error rate controlling or estimating procedures address the multiple comparisons from testing many categories.

---

**3. Class 1 gene category tests.** Global test statistics detect an increased level of DE among the genes within a category. Many testing procedures use traditional methods for comparing independent samples from two populations. In the proposals for these methods, the null hypothesis is rarely stated, and without discussion of the appropriateness of the underlying assumptions. While a variety of global statistics have been employed in these tests, and $p$-values are obtained from both exact and approximate distributional assumptions, we note the null hypotheses have a common form.

DEFINITION 1. A gene category test is of Class 1 if it assumes (or induces through gene permutation) the null hypothesis that the local statistics

$T_1, \ldots, T_m$ are independent and identically distributed (i.i.d.), namely,

(3.1) \qquad $H_0 : T_1, T_2, \ldots, T_m$ are i.i.d. with $T_i \sim F$,

where $F$ can take any form.

3.1. *A survey of global test statistics.* The global statistics proposed for Class 1 tests fall into two groups. "Categorical" statistics rely on a list of significant genes to have been identified by a prior gene-specific analysis, while "continuous" global statistics incorporate real-valued measures of DE for each gene, without reference to a list of significant genes. To illustrate the variety of global statistics that have been proposed in the literature, we present two examples from each group and give a brief description of the corresponding nonresampling based Class 1 tests. A one-sided form of each test is given, because in most applications one is only interested in categories showing more association with the response than their complements. We note it is conceivable to conduct a one-sided test in the opposite direction; for instance, one could look for relative stability within a set of housekeeping genes.

*Categorical statistics.* Gene-list enrichment methods have been developed as a post hoc means of testing a category once the genes with significant DE have been identified. Let $\Gamma$ denote the rejection region for local statistics that produces the list of significant genes. Categorical methods consider only the dichotomous outcomes of the $m$ gene-specific hypothesis tests, and the extent of DE within $C$ and $\bar{C}$ can therefore be summarized by a $2 \times 2$ contingency table [illustrated in Supplementary Figure 1, Barry, Nobel and Wright (2008)].

Traditional tests for contingency tables have been utilized in various gene category analyses, including the $\chi^2$ test of homogeneity, Fisher's Exact test, and slight variations on these tests for contingency tables. In the classical derivation of such tests, binary variables $I\{T_1 \in \Gamma\}, \ldots, I\{T_m \in \Gamma\}$ are assumed to be independent with probabilities of rejection $P(T_i \in \Gamma) = \pi_C$ for $i \in C$ and $P(T_i \in \Gamma) = \pi_{\bar{C}}$ for $i \in \bar{C}$. The tests are designed to have power to detect departures from (3.1) of the form $\pi_C \geq \pi_{\bar{C}}$ under the assumption that the indicator variables are i.i.d. It is worthwhile to note that the Class 1 null is sufficient, but not necessary, for the dichotomous outcomes to be i.i.d. under a given $\Gamma$. However, (3.1) guarantees the categorical null holds for any possible choice of rejection region. We also note that $\Gamma$ may at times be defined in a data-dependent manner, such as when using an error controlling procedure in defining the significant gene list. This violates the assumption of independent test results, even if expression is uncorrelated between genes.

The most common test in gene-list enrichment methods is Fisher's Exact Test. Formally, this is a conditional test based on the total number of rejected hypotheses, $R = \sum_{i=1}^{m} I\{T_i \in \Gamma\}$. The global statistic can be represented as the number of genes in the category that are rejected, namely,

$$(3.2) \qquad U_F = \sum_{i \in C} I\{T_i \in \Gamma\}.$$

Given $R$, an exact $p$-value can be obtained from the hypergeometric distribution.

In several gene-list enrichment software packages, the unconditional $\chi^2$ test of homogeneity is proposed as an approximate test for large categories [Draghici et al. (2003) and Beißbarth and Speed (2004)]. The one-sided version of this test is equivalent to the difference in proportions test originally proposed by Pearson (1911). The associated global statistic can be written in the form

$$(3.3) \quad U_P = \frac{\hat{\pi}_C - \hat{\pi}_{\bar{C}}}{\hat{\sigma}_P} = \frac{1}{m_C \cdot \hat{\sigma}_P} \sum_{i \in C} I\{T_i \in \Gamma\} - \frac{1}{m_{\bar{C}} \cdot \hat{\sigma}_P} \sum_{i' \in \bar{C}} I\{T_{i'} \in \Gamma\},$$

where $\hat{\sigma}_P$ is the traditional estimated standard deviation of the difference in proportions. Under the Class 1 null, the central limit theorem ensures that the two proportions are asymptotically normal for large $m_C$ and $m_{\bar{C}}$, such than a $Z$-test can be performed on $U_P$.

Given the variety of methods for generating gene-lists, it is not always clear whether it is appropriate to condition on $R$, but in general, exact tests are favored for their ability to handle small categories. For moderately sized categories, we note there will be little difference between the exact conditional and the following approximate unconditional test. For this reason, we will restrict our attention to $U_F$ in the simulations performed in Section 4.

*Continuous statistics.* In contrast to gene-list type tests, it is also possible to directly compare the observed associations of expression and response without an intermediate gene list. One straightforward global statistic is the average difference in local statistics between a category and complement, namely,

$$(3.4) \qquad U_D = \frac{\hat{\mu}_C - \hat{\mu}_{\bar{C}}}{\hat{\sigma}_D} = \frac{1}{m_C \cdot \hat{\sigma}_D} \sum_{i \in C} T_i - \frac{1}{m_{\bar{C}} \cdot \hat{\sigma}_D} \sum_{i' \in \bar{C}} T_{i'},$$

which has power to detect an increase in the expected value of local statistics in the category, $\mu_C = E[T_i | i \in C]$, relative to the complement, $\mu_{\bar{C}} = E[T_i | i \in \bar{C}]$ [as illustrated in Supplementary Figure 1, Barry, Nobel and Wright (2008)]. Several hypothesis tests based on the average difference have been proposed, including a $Z$-test performed where $\hat{\sigma}_D$ is the standard deviation of all $m$ local statistics [Kim and Volsky (2005)], and a $t$-test performed where $\hat{\sigma}_D$ is the pooled sample variance of the local statistics [Boorsma et al. (2005)]. In the remainder of this paper we will focus on the $t$-test version of $U_D$, but

note for a typical category where $m_C \ll m$, the variance estimates in the two approaches will be similar, yielding comparable results.

The global statistic in (3.4) may not be robust to outliers or skewness in the local statistics. Rank-based global statistics avoid this shortcoming, as they are invariant to monotone transformations of the local statistics. The Wilcoxon rank sum statistic,

$$U_W = \sum_{i \in C} \text{Rank}(T_i), \tag{3.5}$$

is designed to test a median difference in the two populations of local statistics and has been implemented in GOStat by Beißbarth and Speed (2004). Under the Class 1 null hypothesis, the discrete CDF of $U_W$ is known once $m_C$ and $m_{\bar{C}}$ are specified. Hypothesis testing then proceeds using an exact procedure or a normal approximation to $U_W$.

A Kolmogorov–Smirnov type global statistic has also been implemented in another rank-based Class 1 procedure [Ben-shaul, Bergman and Soreq (2005)]. However, the Kolmogorov–Smirnov statistic has been criticized in gene category testing for being sensitive to departures that do not necessarily reflect increasing DE in the category [Damian and Gorfine (2004)]. For example, a category with no DE but with local statistics that all happen to be nearly identical would be considered significant by these tests. For this reason, we restrict our focus to $U_D$ and $U_W$ when considering continuous global statistics.

3.2. *Gene permutation.* Several permutation-based methods have proposed randomly reordering the rows of the data matrix to determine category significance [Ashburner et al. (2000), Pavlidis et al. (2004) and Zhong et al. (2004)]. In this setup, the collection of local statistics remains unchanged while the category assignments are randomized. Gene permutation effectively induces the Class 1 null hypothesis in (3.1), with the distribution of each reassigned local statistic equaling the empirical distribution of all observed values, $\hat{F}(t) = m^{-1} \sum_{i=1}^{m} I\{t_i \le t\}$. Exhaustive permutation of the gene assignments will be identical to a Fisher's Exact Test of $U_F$ and a Wilcoxon rank sum test of $U_W$. Although gene permutation has limited usefulness for global statistics with traditional tests for the null stated in (3.1), it has proven to be useful in more complex global statistics [Efron and Tibshirani (2007)], and also maintains the correlation between tests of overlapping categories, despite inducing independent local statistics.

We emphasize that gene-permutation procedures (which are also called "gene-shuffling") follow a reasonable and principled development, if one is willing to assume the category assignments in $C$ are random. Then the null hypothesis is that $C$ and the expression data $X$ are independent. Under these assumptions, gene permutation reflects inference conditioned on the expression data. However, as we detail later, Class 1 tests, including those based on gene permutation, are sensitive to correlation of expression of genes within categories, regardless of DE. Such correlation represents a departure from the assumption of independence of $X$ and $C$, but is unrelated to DE. Following our perspective that gene category assignments are fixed, we view gene permutation procedures as Class 1.

**4. The effect of correlation on Class 1 tests.** In this section we examine more closely the assumption of independent local statistics, and how violations of this assumption effect the performance of Class 1 tests. We note that correlation of local statistics arises naturally from correlation of expression among genes. A simulation study based on a real microarray data set exhibits the extreme anti-conservative behavior of Class 1 tests in the presence of realistic levels of correlation in expression.

4.1. *Correlations in expression and local statistics.* Let the population correlation between genes $i$ and $i'$ be given as $\rho^X_{i,i'} = \text{Corr}(X_{ij}, X_{i'j})$. For experimental designs with independent arrays, a natural estimate of $\rho^X_{i,i'}$ is the sample correlation coefficient, $r_{i,i'}$. The distributions of global statistics for Class 1 tests are directly affected by the correlation between local statistics, $\rho^T_{i,i'} = \text{Corr}(T_i, T_{i'})$. In the special case that $T$ takes the linear form $T(\mathbf{X}_{i*}, \mathbf{y}) = \sum_{j=1}^n a(y_j) \cdot X_{ij}$ for some function $a(\cdot)$, it is easy to see that $\rho^T_{i,i'} = \rho^X_{i,i'}$. An example of a linear local statistic would be fold change on the log-scale.

In general, the relationship between $\rho^X_{i,i'}$ and $\rho^T_{i,i'}$ does not have a simple analytic form, although it can be shown numerically to often be monotone and approximately linear for one-sided local statistics. Indeed, Monte Carlo simulations of gene expression data (Figure 1) demonstrate that a nearly linear relationship holds for several standard experimental designs and corresponding measures of DE. This includes using a Student's $t$ as the local statistic for a two-condition study, and a Wald-type statistic for regressing expression on censored time-to-event data through a Cox proportional hazards model. For such local statistics, $\rho^X_{i,i'} \approx \rho^T_{i,i'}$ so that the sample correlation coefficients of gene expression can be used as estimates of $\{\rho^T_{i,i'}\}$ in determining the properties of global statistics. However, the two correlations have a nonlinear relationship for "undirected" local statistics, such as an analysis of variance $F$-statistic [Figure 1(b)].

4.2. *Variance inflation.* The effects of pairwise correlation on Class 1 tests can be illustrated by deriving the true variances of the global statistics $U_D$ and $U_W$ in the presence of dependence.

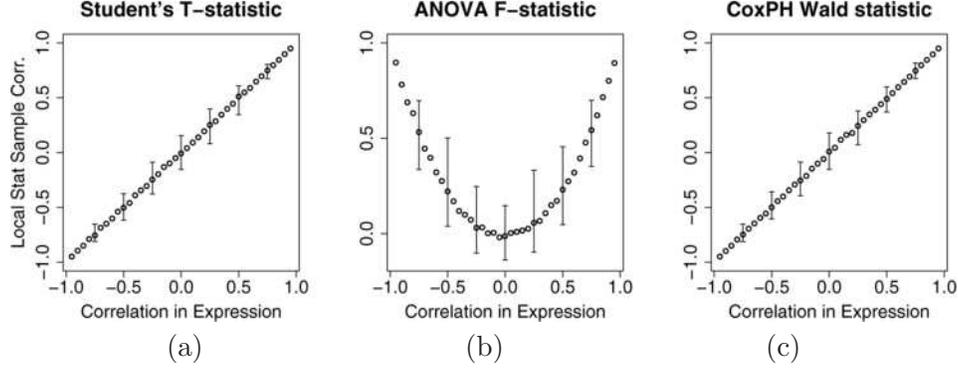

FIG. 1. *Correlations between gene expression levels induce correlations in local statistics. Monte Carlo simulations of standard Gaussian expression for two genes under several experimental designs:* (a) *a two-sample comparison with a Student's t-statistic;* (b) *four-sample comparison with an ANOVA F-statistic;* (c) *survival using a Wald test from a univariate Cox-proportional hazard model. For each design, sample correlation was observed across 100 gene pairs and* $n = 40$ *arrays with equally sized groups or exponentially distributed event and censor times. The median and representative (0.05, 0.95) quantile intervals are shown from 200 simulations. Similar results are obtained when simulating heteroscedastic genes.*

For the average difference global statistic $U_D$, a simple calculation shows that the true variance of the statistic will differ from that under the i.i.d. null in Class 1 tests by three additional terms:

$$\begin{aligned}
&\text{Var}[\hat{\mu}_C - \hat{\mu}_{\bar{C}}] \\
(4.1) \quad &= \text{Var}_{\text{i.i.d.}}[\hat{\mu}_C - \hat{\mu}_{\bar{C}}] \\
&\quad \times \left(1 + \frac{m_{\bar{C}}(m_C - 1)}{m}\rho_C + \frac{m_C(m_{\bar{C}} - 1)}{m}\rho_{\bar{C}} - \frac{m_C \cdot m_{\bar{C}}}{m}\rho_{C,\bar{C}}\right),
\end{aligned}$$

where the quantities

$$(4.2) \quad \rho_C = \frac{1}{m_C \cdot (m_C - 1)} \sum_{i \in C} \sum_{\substack{i' \in C \\ i' \neq i}} \rho^T_{i,i'},$$

$$(4.3) \quad \rho_{\bar{C}} = \frac{1}{m_{\bar{C}} \cdot (m_{\bar{C}} - 1)} \sum_{i \notin C} \sum_{\substack{i' \notin C \\ i' \neq i}} \rho^T_{i,i'},$$

$$(4.4) \quad \rho_{C,\bar{C}} = \frac{1}{m_C \cdot m_{\bar{C}}} \sum_{i \in C} \sum_{i' \notin C} \rho^T_{i,i'}$$

are related to the average pairwise correlations within the category (4.2), within its complement (4.3), and across the two gene sets (4.4). We note that $\rho_C$ can vary greatly across categories, while $\rho_{\bar{C}}$ and $\rho_{C,\bar{C}}$ will be close to the average pairwise correlation of all genes on the array and near zero in most datasets. For a moderately sized category where $m_{\bar{C}} \approx m$, the ration of variances in (4.1), $\text{Var}[\hat{\mu}_C - \hat{\mu}_{\bar{C}}]/\text{Var}_{\text{i.i.d.}}[\hat{\mu}_C - \hat{\mu}_{\bar{C}}]$, is approximately $1 + (m_C - 1) \cdot \rho_C$. This ratio measures the variance inflation of $U_D$ over what is assumed by (3.1), and as a consequence, the category exhibiting positive correlation will tend to have anti-conservative Class 1 tests of significance DE.

For the Wilcoxon rank sum global statistic, the true variance will depend on the common distribution $F$ of local statistics, as defined in (3.1). In the special case that local statistics are marginally normally distributed, with common mean, unit variances, and pairwise correlations $\{\rho_{i,i'}^T\}$, then $\text{Var}[U_W]$ is given by

$$(4.5) \quad \text{Var}[U_W] = \frac{1}{2\pi} \sum_{i \in C} \sum_{i' \in C} \sum_{h \notin C} \sum_{h' \notin C} \sin^{-1}\left(\frac{\rho_{i,i'}^T + \rho_{h,h'}^T - \rho_{i',h}^T - \rho_{i,h'}^T}{\sqrt{(2 - 2\rho_{i,h}^T) \cdot (2 - 2\rho_{i',h'}^T)}}\right).$$

The derivation of (4.5) is provided in the Supplementary material [Barry, Nobel and Wright (2008)] and is analogous to the classic work of Gastwirth and Rubin (1971) on the effect of dependence on the Wilcoxon rank sum. If the local statistics within a category were all positively correlated and the complementary set of genes were independent, this variance is easily shown to be strictly greater than what is assumed under the Class 1 null, $\text{Var}_{\text{i.i.d.}}[U_W] = m_C \cdot m_{\bar{C}} \cdot (m + 1)/12$.

Correlation between local statistics will also affect the distributions of $U_F$ and $U_P$. However, the variance of these categorical global statistics in the presence of correlation will further depend on both the distribution $F$ and the rejection region $\Gamma$. In the next subsection we present a simple simulation study illustrating the effect of gene correlation on Class 1 tests for annotated categories in a real microarray dataset. The anti-conservative behavior of a Class 1 test using the global statistic $U_F$ is explored in the recent work of Goeman and Buhlmann (2007), who considered simulated Gaussian expression data in which the pairwise correlation of genes is fixed and equal, and categories are of a fixed size. The simulation study below attempts to capture the more complicated correlation structures and variable sizes of functional categories that occur in real data.

4.3. *A simulation study.* A two-condition experiment was simulated using a subset of the lung carcinoma microarrays from Bhattacharjee et al. (2001). We first selected 100 adenocarcinoma samples at random from the dataset, that contains expression estimates for 7299 genes [see Barry, Nobel and Wright (2005) for data pre-processing steps]. Among the available genes, 1823 GO and Pfam categories were identified with at least 5 members. The within-category average pairwise sample correlations ranged from −0.09 to 0.93, with more than 86% of the categories having values greater than the average pairwise correlation across the entire array (0.012). This increase in correlation within categories illustrates the common observation that coexpression among genes is associated with gene function [Lee et al. (2004)].

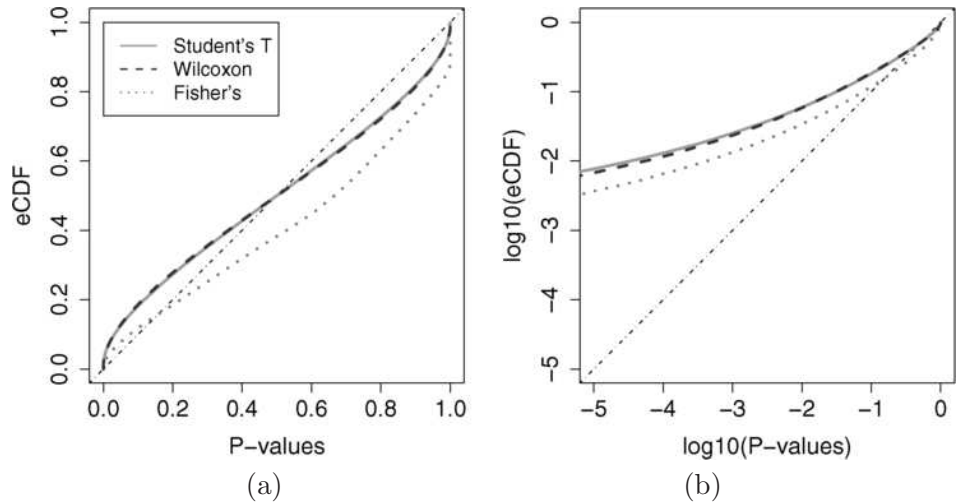

Fig. 2. *Poor performance of Class 1 tests.* (a) *Empirical CDFs of pooled p-values (1823 categories over 1000 simulations) for the Class 1 Fisher's Exact Test, Student's t-test, and Wilcoxon rank sum test generated under the null hypothesis in* (5.1). (b) *The same data are plotted on the log-10 scale to demonstrate the disproportionate number of small p-values that results from incorrect variance estimates.*

One thousand binary response vectors with equal numbers of zeros and ones were generated at random, and used to assign the arrays to one of two conditions. The random assignment of arrays ensured that the resulting datasets had no association between expression and experimental condition for any gene, and thus, no category is expected to have a greater degree of DE than any other. We note that the expression matrix is held constant across simulations, so that the sample gene–gene correlations $\{r_{i,i'}\}$ remained fixed.

For each realization of the response vector, a pooled-variance $t$-statistic was used as the local statistic, and global statistics $U_F$, $U_D$ and $U_W$ were computed. For the Fisher's Exact Test statistic, $U_F$, the rejection region was equal to values exceeding $t_{98,0.95} = 1.66$. For each global statistic and each category, the different Class 1 tests produced a nominal $p$-value for every generated response vector. Empirical CDFs of the nominal $p$-values pooled across all categories and all realizations demonstrate their extreme nonuni-formity under the induced null hypothesis, confirming the poor performance of Class 1 tests (Figure 2).

The average Type I error of these tests was estimated as the proportion of $p$-values under simulations that fall below a target $\alpha$ level. Table 1 displays that, for each global statistic, the corresponding Class 1 tests reject far more hypotheses than expected. Moreover, the anti-conservative behavior of these tests increases for smaller target $\alpha$ values. Even though the gene-list enrichment methods are slightly less anti-conservative than the continuous methods, this is offset by their potential loss in power from dichotomizing local statistics.

To illustrate how this behavior also affects the family-wise error rate among the $L = 1823$ categories, we applied a Bonferroni correction to the nominal $p$-values. Since for the randomized data all categories are truly null, the FWER is estimated by

$$(4.6) \qquad FWER = \frac{1}{1000} \sum_{b=1}^{1000} I\left\{ \sum_{h=1}^{L} I\left\{ p_{b,h} < \frac{\alpha}{L} \right\} > 0 \right\},$$

where $p_{b,h}$ is the Class 1 $p$-value for category $h$ under realization $b$. There is substantial overlap in the membership of gene categories from annotations such as Gene Ontology, and thus tests will be positively correlated accordingly. Therefore, the use of Bonferroni thresholds might be thought to be overly stringent in controlling the FWER, providing some protection against anti-conservative Class 1 $p$-values. However, for $\alpha = 0.05$, the realized FWER in (4.6) is far greater than the target level ($U_F : 0.776$, $U_D : 0.925$ and $U_W : 0.918$). The extreme anti-conservative behavior of the Class 1 tests of all global statistics suggests a different approach is needed to conduct valid gene category tests.

**5. Class 2 tests and array permutation.** The null hypothesis of Class 1 tests is violated by the correlations present in gene expression data, and we demonstrate the resulting anti-conservative behavior. For this reason, a second class of gene category tests is warranted that can identify increases in DE within a category, while properly accounting for correlation.

DEFINITION 2. Class 2 gene category tests are defined by the assumed or induced null hypothesis that the local statistics $T_1, \ldots, T_m$ are possibly dependent and identically distributed. More precisely,

$$(5.1) \quad H_0 : T_1, T_2, \ldots, T_m \text{ are identically distributed with } T_i \sim F_0,$$

where $F_0$ corresponds to a lack of association between expression and the response of interest. No assumptions are made about dependence among the $T_i$s.

In order for (5.1) to hold in a given experimental design, an appropriate form of $T(\cdot)$ must be selected to ensure local statistics are marginally identically distributed. In the following section we describe a sufficient property of local statistics to induce (5.1) under a global null of all genes having no DE.

5.1. *$\delta$-determined local statistics.* In gene category testing, the true association between the expression of an individual gene and a real- or vector-valued response can often be summarized by a single fixed parameter $\delta$ that depends on their (unknown) joint distribution. In the absence of any association, the parameter frequently assumes a known null-value. Accordingly, the local statistic $T(\cdot)$ is chosen for its utility in conducting hypothesis tests against the null value of the parameter. To illustrate, consider a two-condition experiment where the response vector **y** takes values of 1 or 2, indicating the sample condition of each array. If the expression of gene $i$ has expectation $\mu_{1i}$ and $\mu_{2i}$ under the two respective conditions, and common variance $\sigma_i^2$, then a natural measure of association is the scaled difference in means

$$(5.2) \qquad \delta_i = \frac{\mu_{1i} - \mu_{2i}}{\sigma_i \cdot \sqrt{1/n_1 + 1/n_2}}.$$

Here and below, $\delta_i$ denotes the value of the association parameter for gene $i$. In this case, the gene-specific null hypothesis of interest is $H_{0,i} : \delta_i = 0$, and the pooled-variance $t$-statistic is a natural choice of local statistic [Galitski et al. (1999)]. When the expression levels of gene $i$ are normally distributed, and independent from sample to sample, the local statistic follows a $t$-distribution with noncentrality parameter $\delta_i$, which reduces to a central $t$-distribution when $\delta_i = 0$.

In general, a function $T(\cdot)$ is a proper choice of test statistic for a null of the form $H_0 : \delta = d$ when the distribution $F(T|\delta_i = d)$ of $T$ given $\delta = d$ is known and does not depend on any nuisance parameters. When the distribution of $T(\cdot)$ can be specified in this manner for any choice of $d$, we refer to its distribution as being *$\delta$-determined*. This property is important in the basic theory of interval estimation and pivotal quantities; if $F(T|\delta = d)$ is $\delta$-determined, it can be used as a pivotal quantity to construct a confidence set for $\delta$ [Casella and Berger (2002)]. In the particular example presented above, a Student's $t$ is $\delta$-determined by (5.2).

The $\delta$-determined property is important when conducting gene category tests; if the distribution of $T$ is $\delta$-determined, then differences in nuisance parameters do not influence the comparison of a category against its complement. To illustrate, consider a two-condition experiment with $\delta$ defined as (5.2). Here the gene-specific means and variances of expression are considered nuisance parameters. Suppose that for each gene one directly uses the modified $t$-statistic from the SAM software [Tusher, Tibshirani and Chu (2001)] as the local statistic. This statistic contains a constant in the denominator that effectively penalizes lowly-expressed genes in order to improve the FDR of a gene-list. Due to the presence of this constant, the SAM $t$-statistic is not $\delta$-determined, as its distribution depends on the means and variances of gene expression. Now consider a category consisting primarily of highly-expressed housekeeping genes. Under a global null in which no genes are differentially expressed, and thus no category should be considered significant, highly expressed genes have an increased chance of being ranked above lowly-expressed genes when using the SAM statistic. Thus, a category with highly expressed genes is more likely to be (falsely) declared significant, regardless of whether one uses a categorical or continuous global statistic. Categories with lowly-expressed genes would experience the opposite effect.

When a $\delta$-determined local statistic is chosen and a unique value, $d_0$, of the parameter corresponds to a lack of association between expression and response, the Class 2 null can be restated as $H_0 : \delta_1 = \cdots = \delta_m = d_0$. For the remainder of the paper, we will only consider local statistics that are $\delta$-determined, or approximately so when $n$ is large.

5.2. *Array permutation.* If the pairwise correlations $\{\rho_{i,i'}^T\}$ of the local statistics are known, a Class 2 test can be constructed for the average difference statistic $U_D$ using its true variance, as derived in (4.1). Similarly, an approximate $Z$-test for the Wilcoxon rank sum statistic $U_W$ can be designed using (4.5) if local statistics are approximately normal. However, since the correlations are generally unknown, a particular form of permutation can be used as an alternative means of approximately inducing the Class 2 null.

In many common microarray experiments, each mRNA sample constitutes an independent unit. By permuting the column vectors of $\mathbf{X}$, or equivalently the response vector $\mathbf{y}$, an empirical null distribution is achieved in which there is no association between gene expression and the response. Array permutation was first used in Virtaneva et al. (2001) to test categories of genes, and then implemented in GSEA for a Kolmogorov–Smirnov global statistic [Mootha et al. (2003)], and in SAFE for any chosen global statistic [Barry, Nobel and Wright (2005)]. More recently, other global statistics have been proposed for use with array permutation, including a weighted version of GSEA [Subramanian et al. (2005)] and a standardized truncated mean that is more sensitive to directional changes [Efron and Tibshirani (2007)]. These reports note that array permutation does not change the correlations in expression among genes, and thus the gene-specific measures of DE remain dependent. Also, when using array permutation the resampled local statistics are conditional on the observed dataset, such that their empirical distributions are not exactly identically distributed, and only approximately follow the Class 2 null. However, if one uses the gene-specific empirical $p$-values as local statistics, then every local statistic will exactly follow the discrete uniform distribution under permutation, guaranteeing (5.1).

5.3. *Simulated coverage of Class* 2 *tests.* The randomized datasets from (4.3) were used to evaluate Class 2 tests of each global statistic based on array permutation. Here the tests are ensured to be of proper size, since the randomization procedure in the simulation and the array permutation in the test both employ the same sampling process. We confirmed this by obtaining empirical $p$-values for each category and each realization of the response vector (Table 1). Due to the computational burdens of both simulation and permutation, only 1000 resamples were taken for each test, so the smallest possible empirical $p$-value was 0.001. The Class 2 Fisher's Exact Test results are notably conservative, due to the numerous tied global statistics that occur in small categories. The slight misspecification of Type I error for $U_P$, $U_D$ and $U_W$ reflects only sampling variability. These results demonstrate Class 2 tests of gene categories generally outperform Class 1 tests based on the assumption of gene independence.

**6. A more general null for gene category tests.** Although Class 2 tests of gene categories appropriately account for the correlation structure of gene expression data, they share with Class 1 procedures the shortcoming of assuming a null hypothesis under which local statistics are identically distributed. This assumption is not necessary when considering whether a gene category exhibits a greater degree of differential expression than its complement. For example, suppose 20% of the genes are differentially expressed to the same degree (where $\delta_i$ equals a common nonnull value $d$), and the remaining genes have no association with the response ($\delta_i$ equals the null value $d_0$). Any category in which 20% of the genes are differentially expressed should not be considered "special." However, the Class 2 null, which is induced by array permutation, is clearly violated under this scenario.

TABLE 1
*The ratio of realized Type* I *error rates to target* $\alpha$ *levels*

|  | Fisher's, $U_F$ | Student's $t$, $U_D$ | Wilcoxon, $U_W$ |
|---|---|---|---|
| Class 1 tests |  |  |  |
| $\alpha = 0.1$ | 1.19 | 1.82 | 1.86 |
| $\alpha = 0.01$ | 3.40 | 5.92 | 5.83 |
| $\alpha = 0.001$ | 13.4 | 25.2 | 23.5 |
| $\alpha = 1e\text{--}4$ | 65.6 | 130 | 116 |
| $\alpha = 1e\text{--}5$ | 367 | 769 | 677 |
| $\alpha = 1e\text{--}6$ | 2213 | 4974 | 4245 |
| Class 2 tests |  |  |  |
| $\alpha = 0.1$ | 0.39 | 1.01 | 1.01 |
| $\alpha = 0.01$ | 0.21 | 1.01 | 1.01 |
| $\alpha = 0.001$ | 0.14 | 1.03 | 1.01 |
| $\alpha = 1e\text{--}4$ | NA | NA | NA |

Based on this simple example, we propose the following less restrictive, and more biologically realistic, null hypothesis of gene categories. Instead of requiring the local statistics of all genes to be identically distributed, we allow each to fall into one of $K \leq m_C$ strata that correspond to a different marginal distribution for the statistic. No conditions are imposed on the dependence of the local statistics. The null can be formalized as follows.

DEFINITION 3. Let $K \geq 1$ and let $G_1, \ldots, G_K$ be distinct, fixed distributions. Let the local statistics $T_1, \ldots, T_m$ have marginal distributions $F_1, \ldots, F_m$ and let $C \subset \{1, \ldots, m\}$ be a gene category. Assume that each $F_i \in \{G_1, \ldots, G_K\}$ and that $\beta_{C,k} = m_C^{-1} \sum_{i \in C} I\{F_i = G_k\}$ and $\beta_{\bar{C},k} = m_{\bar{C}}^{-1} \sum_{i \in \bar{C}} I\{F_i = G_k\}$ are the proportions of genes from $C$ and $\bar{C}$, respectively, whose local statistics are distributed as $G_k$. The Class 3 null hypothesis is the following:

(6.1) $$H_0 : \beta_{C,k} = \beta_{\bar{C},k} \equiv \beta_k, \qquad k = 1, \ldots, K,$$

where the distributions $G_1, \ldots, G_K$ can take any form.

The Class 1 and Class 2 null hypotheses are then special cases of (6.1) with $K = 1$ stratum. When DE can be assessed in terms of an association parameter $\delta$, and the local statistic is $\delta$-determined [e.g., the scaled difference in means (5.2) and the Student's $t$-statistic], the strata can be directly related to different degrees of association with the response. In this case, the Class 3 null hypothesis is equivalent to the statement that the empirical distributions of $\delta$s in $C$ and $\bar{C}$ are identical.

Dudoit et al. (2007) apply a general approach to multiple testing to a family of problems involving simultaneous testing of annotation-based profiles using gene expression data. Their work provides a framework for multiple testing of associations between what the authors term gene-annotation profiles and gene-parameter profiles. The former include gene categories and GO terms. The latter are population based quantities that encompass measures of association between the expression of a gene and a binary or continuous response, including a scaled difference (or ratio) of means. When local statistics are $\delta$-determined, the Class 2 and Class 3 nulls considered here can be placed within that framework, with the vector of gene-parameter profiles playing a role analogous to the gene association parameters $\delta$. We note that the Wilcoxon global statistic is not linear in the sense described in Dudoit et al. (2007), and therefore is not covered by their theoretical developments.

In the following subsections we discuss simple bootstrap-based tests that: (a) maintain the correlation structure of the expression data and (b) demonstrate approximate Type I error control under different realizations of the Class 3 null. Simulations of microarray data reveal that tests based on bootstrap resampling of arrays clearly outperform array permutation tests that induce a Class 2 null. Further examination of the distributional properties of $U_W$ and $U_D$ indicate the poor performance of array permutation in this more general setting. Finally, we illustrate through simulation the increased power of Class 3 tests under a defined set of alternative hypotheses.

6.1. *Defining the bootstrap-based tests.* Standard bootstrap methodology assumes that the observed data can be divided into independent units derived from an unknown probability model. Resampling from the empirical distribution of the observed data enables one to form approximate confidence intervals without parametric assumptions [Efron and Tibshirani (1998)]. For most microarray experiments, the independent sampling unit is the joint vector $\{\mathbf{x}_{*j}, y_j\}$ containing $m$ gene expression measurements and response information for a single mRNA sample. To approximate the unknown probability model of the data, bootstrap procedures resample the joint vectors with replacement. Let $\mathbf{b} = (b_1, \ldots, b_n)$ be a resampling vector whose elements are independent and uniformly distributed over the integers $\{1, \ldots, n\}$. Associated with $\mathbf{b}$ is a resampled response $\mathbf{y}^{*b} = (y_{b_1}, \ldots, y_{b_n})$, and a resampled expression matrix in which the measurements of gene $i$ are given by $\mathbf{x}_{i\cdot}^{*b} = (x_{ib_1}, \ldots, x_{ib_n})$. From the resampled data, local statistics $t_i^{*b} = T(\mathbf{x}_{i\cdot}^{*b}, \mathbf{y}^{*b})$, and a global statistic $u^{*b} = U(t_1^{*b}, \ldots, t_m^{*b} : C)$ may be calculated in the usual way. Let $B$ denote the total number of bootstrap samples.

Standard procedures can be used to generate bootstrap confidence intervals for the parameter $\theta = E[U]$. In the context of looking for increased amounts of DE, one would define a one-sided confidence interval of size $\alpha$ by its lower bound $L_\alpha$. Arguably, the simplest procedure for producing the one-sided confidence interval via bootstrap resampling is the quantile method [Efron (1979)], in which $L_\alpha$ is the sample $\alpha$-quantile of the resampled values: $u^*_{(B\cdot\alpha)}$. The quantile method is straightforward to compute and invariant under monotone transformations of the global statistics. However, its coverage may be poor when the sample size is small [Efron (1987)], due to the difficulty of estimating the tail distribution of the global statistic. Alternatively, if one assumes that the global statistic is approximately normal, a confidence interval can be generated from the $t$-distribution using bootstrap-based estimates of the moments of $U_W$ [Efron (1979)]. The resulting one-sided confidence interval has a lower bound given by

$$\bar{u}^* - \widehat{se}^*(U) \cdot t_{n-1,1-\alpha}, \tag{6.2}$$

where

$$\bar{u}^* = \frac{1}{B}\sum_{b=1}^{B} u_b^* \quad \text{and} \quad \widehat{se}^*(U) = \left[\frac{\sum_{b=1}^{B}(u_b^* - \bar{u}^*)^2}{B-1}\right]^{1/2}. \tag{6.3}$$

Hypothesis testing with the bootstrap was originally proposed by Efron as being applicable when the distribution of a test statistic was unknown under a null hypothesis (due to nuisance parameters) but could be induced by bootstrap-resampling [Efron and Tibshirani (1998)]. Hall and Wilson (1991) have proposed that bootstrap-based tests can also be constructed when the empirical distributions of resampled statistics do not directly relate to $H_0$ if a "pivot test" is used. In the setting of gene categories, this would relate to testing for the exclusion of some null value, $\theta_0 = E_{H_0}[U]$, from a confidence interval defined by the populations of resampled global statistics. These tests are of particular use for the Class 3 null, which would be difficult to induce directly through the resampling the arrays in a way that also maintains gene-correlation.

We generally favor using the Wilcoxon rank sum global statistic, $U_W$, for gene category tests, because it is a robust and transformation-invariant measure of average difference that avoids the arbitrariness the gene-list methods have of choosing a rejection region. The following theorem establishes the expectation of the Wilcoxon global statistic $U_W$ is a constant, $\theta_0$, under all realizations that the stratified null hypothesis (6.1).

THEOREM 1. *Suppose that for each $i$ the local statistic $T_i$ of gene $i$ has distribution $F_i \in \{G_1, \ldots, G_K\}$ and that $P(T_i = T_j) = 0$ for $i \neq j$. Then for any category $C \subset \{1, \ldots, m\}$ such that $\beta_{C,k} = \beta_{\bar{C},k} = \beta_k$ for each stratum $k = 1, \ldots, K$,*

$$E[U_W] = \frac{m_C \cdot (m+1)}{2}. \tag{6.4}$$

(See Appendix A for the proof.) Note that the expectation is constant, regardless of the number of strata $K$, the proportions $\{\beta_1, \ldots, \beta_K\}$, and the distributions $\{G_1, \ldots, G_K\}$. Similar derivations demonstrate that the global statistics $U_D$ and $U_P$ have a fixed expectation of 0 under (6.1), allowing for the construction of a test based on bootstrap resampling.

The Class-3 bootstrap tests advocated here are based on standard bootstrap confidence intervals. Dudoit et al. (2007) also propose bootstrap tests within the multiple testing framework considered in their paper. Specialized to the setting of the Class 3 null, their bootstrap procedure is essentially similar to the quantile-based method we discuss above. Dodd et al. (2006) have also applied bootstrap resampling to category testing in the context of differential expression between cancerous and normal human tissue, in which a genelist was interrogated with the Pearson-type global statistic, $U_P$. Many of the arguments we present in the following sections in support of bootstrap testing can be applied to the procedures in Dudoit et al. (2007) and Dodd et al. (2006).

In contrast to the global statistics mentioned above, the expectation of the global statistic employed in Fisher's Exact test depends on the $K$ gene-specific distributions, and the expectation of the Kolmogorov–Smirnov type global statistic used in Mootha et al. (2003) depends on both the marginal distribution of the local statistics, and their correlation structure. As a consequence, one cannot test for exclusion of a null value from confidence intervals of these two global statistics.

The following simulations of various Class 3 nulls were designed to evaluate the Type I and II errors of bootstrap-based tests for both continuous global statistics, $U_W$ and $U_D$. These results demonstrate that array permutation is clearly inferior in this setting.

6.2. *Type* I *error under a simulated Class* 3 *null.* The randomized lung cancer dataset described in Section 4.3 is extended to evaluate the Type I error incurred by permutation- and bootstrap-based tests of $U_W$ and $U_D$ under (6.1). A Student's $t$ is used as the local statistic for DE across the simulated conditions; under the assumption the gene expression values approximately follow normality, each distributional strata is determined by the association parameter $\delta$ given in (5.2). We investigated several null hypotheses with $K = 2$ strata of genes relating to no DE ($\delta_i = 0$) and positive DE ($\delta_i = d > 0$). To artificially generate different degrees of DE in a particular gene, the expression values were first standardized to have variance 1; then $d \cdot \sqrt{1/n_1 + 1/n_2}$ was added to those values $x_{ij}$ with condition $y_j = 1$. Simulations were run using three different levels of DE, $d = 1, 3$ and $5$, and also for three proportions of DE, $\beta = 1/5, 1/3$ and $1/2$. For each proportion, we selected a subset of nonoverlapping categories with the property that $\beta \cdot m_C$ is an integer. This resulted in 41 categories being considered for $\beta = 1/5$, 40 categories for $\beta = 1/3$ and 34 categories for $\beta = 1/2$. The selected categories exhibited a wide range of correlation in expression, reflective of that seen across the entire dataset.

For each of 10000 randomizations of tumor status, the permutation- and quantile bootstrap-based hypothesis tests were conducted using 2000 permutations and resamples, respectively. Simulations indicate that $B = 200$ resamples are typically sufficient for the moment estimates in the bootstrap $t$-intervals, and were used accordingly. Type I error was determined by comparing the empirically derived $p$-values to various $\alpha$ levels (Figure 3). For a target $\alpha = 0.05$, the bootstrap Type I error was only slightly inflated, and remained relatively unchanged regardless of $\beta$ and $d$, whereas the Type I error of permutation testing dropped dramatically as either parameter diverged from 0. For $d=3$ and $\beta = 1/3$, the minimum empirical $p$-values obtained under permutation were 0.0055 and 0.001 for $U_W$ and $U_D$ respectively, which are several orders of magnitude higher than what would be expected with proper Type I error control.

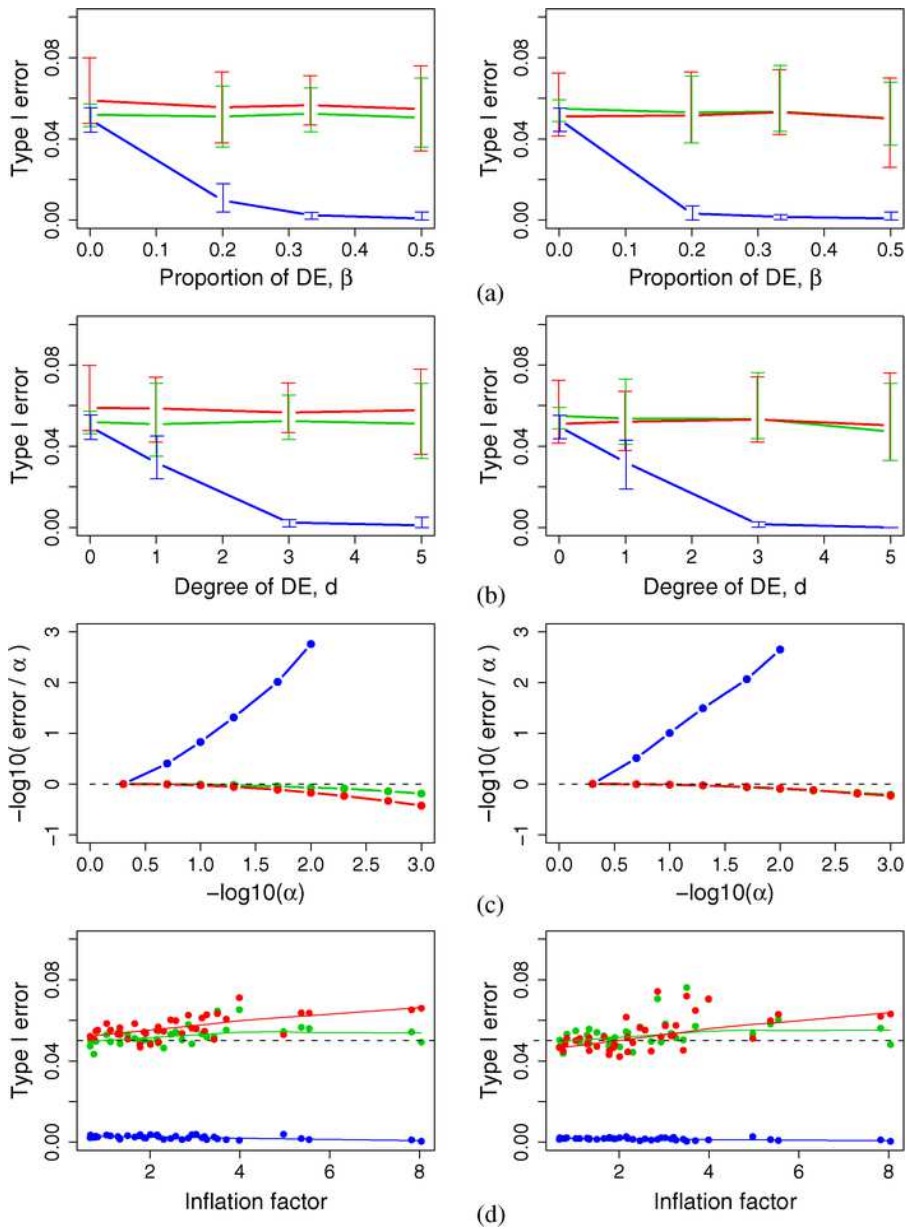

Fig. 3. *Performance of gene category tests using bootstrap-quantile (green), bootstrap-t (red), and array permutation (blue) methods, for the continuous global statistics $U_W$ and $U_D$, and various Class 3 null hypotheses. The range of Type I errors of different categories is shown for* (a) *four proportions of DE, $\beta$, and* (b) *for four levels of DE;* (c) *the average Type I errors of $U_W$ and $U_D$ for different $\alpha$ levels under a Class 3 null with $d = 3$ and $\beta = 1/3$;* (d) *the Type I error for $\alpha = 0.05$ is plotted for each category against their estimated Class 1 variance inflation, $1 + (m_C - 1) \cdot \hat{\rho}_C$, as derived from (4.1).*

These findings illustrate the Class 2 tests based on array permutation are overly conservative under the broader null. In order to better understand the conservative behavior of array permutation, we note that it induces a special case of (6.1) under which the local statistics are approximately identically distributed (5.1). We return to the variance of the Wilcoxon global statistic derived in (4.5), and define the following type of positively correlated category.

DEFINITION 4. For local statistics $T_1, \ldots, T_m$ with correlations $\{\rho^T_{i,i'}\}$, a category $C \subseteq \{1, \ldots, m\}$ will be called *correlation dominant* if for every $\{i, i'\} \in C$ and $\{h, h'\} \notin C$ it is true that $\rho^T_{i,i'} \geq \rho^T_{i,h}$ and $\rho^T_{i,h} \leq \rho^T_{h,h'}$, in other words, correlations within the category and its complement are greater than those between the category and its complement.

In the following theorem we establish that, for normally distributed local statistics, the variance of the Wilcoxon global statistic $U_W$ is maximized under the $K = 1$ null given in (5.1) for all correlation dominant categories.

THEOREM 2. *Let $T_1, \ldots, T_m$ be random variables that follow a multivariate normal distribution with means $\delta_1, \ldots, \delta_m$, unit variances and correlations $\{\rho^T_{i,i'}\}$. For a correlation dominant gene category $C$, the variance of $U_W$ has a global maximum at $\delta_1 = \delta_2 = \cdots = \delta_m = d$.*

Because array permutation induces the special case of (6.1) where the variance of $U_W$ is maximum, it is reasonable that the tests will tend to become conservative under Class 3 null hypothesis that depart from (5.1). Although the complex structure of gene correlation would likely prevent any real category from meeting the strict criterion of being correlation dominant, the conservativeness of array permutation tests is seen across all categories in the randomized datasets (Figure 3). We have also confirmed these results in two-condition datasets simulated from a multivariate Gaussian model (not shown). The equal conservativeness of $U_D$ remains to be explored in full detail, but we suggest that the pooled-variance estimate used in standardizing the global statistics similarly overestimates the true variance in the Class 3 null hypotheses with more than 1 stratum of DE.

While in the simulations presented above both bootstrap methods maintained their approximately correct Type I error throughout, we have found that in simulated expression data with a smaller sample size of $n = 20$ arrays, the quantile-based method becomes more anti-conservative at small target $\alpha$, since many microarray datasets can be of this size, and the quantile-interval also requires more resamples to conduct tests with small $\alpha$. Further, as $U_W$ is the sum of $m_C \cdot (m - m_C)$ pairwise comparisons of local statistics, approximate normality follows from the Central Limit Theorem when the average correlation between these terms is not extreme and $m_C$ is large. Histograms of resampled global statistics confirm that the approximation to the normal distribution is appropriate for the large number of genes in a typical microarray experiment. Therefore, we prefer the bootstrap Student's $t$-interval for more general use in Class 3 tests of gene categories.

6.3. *Power under simulated alternatives.* To assess the relative power of the bootstrap tests over array permutation, alternative hypotheses must be specified that relate to increased amounts of DE in a gene category. When the differential expression of a gene can be measured in terms of an association parameter $\delta$, an average increase of DE within the category relative to its complement can be written as

$$(6.5) \qquad H_A : \sum_{i=1}^{K} \beta_{C,k} \cdot d_k > \sum_{i=1}^{K} \beta_{\bar{C},k} \cdot d_k.$$

For these alternatives, the Wilcoxon rank sum $U_W$ is well suited to identifying increased amounts of DE in a robust manner.

In the randomized lung carcinoma dataset, realizations of (6.5) can be achieved by applying an additive or multiplicative constant to all gene-specific parameters within the category. More precisely, if $\{\delta_i^0 : i \in C\}$ are the association parameters of a category under the Class 3 null, we consider $H_A$ to be either of the form $\{\delta_i^A = c + \delta_i^0 : i \in C\}$ or $\{\delta_i^A = c \cdot \delta_i^0 : i \in C\}$. In this way, power curves can be displayed across a single axis by varying $c$. Figure 4 illustrates the effects when the constant is applied in an additive manner for $K = 2$ strata of DE and nonDE genes, and in a multiplicative manner for an example with $K = 5$ strata. The results demonstrate considerable improvements in power of the bootstrap methods over array permutation.

**7. Analysis of a survival microarray dataset.** The breast cancer survival dataset from Chang et al. (2005) are used to illustrate the utility of bootstrap resampling as compared to array permutation. A total of $n = 295$ breast cancer samples were analyzed on Agilent microarrays, and normalized gene expression estimates were obtained for a subset of $m = 11176$ genes that are annotated to at least one of 1348 GO terms (details on normalization, filtering, and formation of gene categories are omitted, but available from the authors). Survival times and censoring indicators were available for each array. Wald statistics from univariate Cox proportional hazard models were used as local statistics to reflect the association between expression and patient outcome.

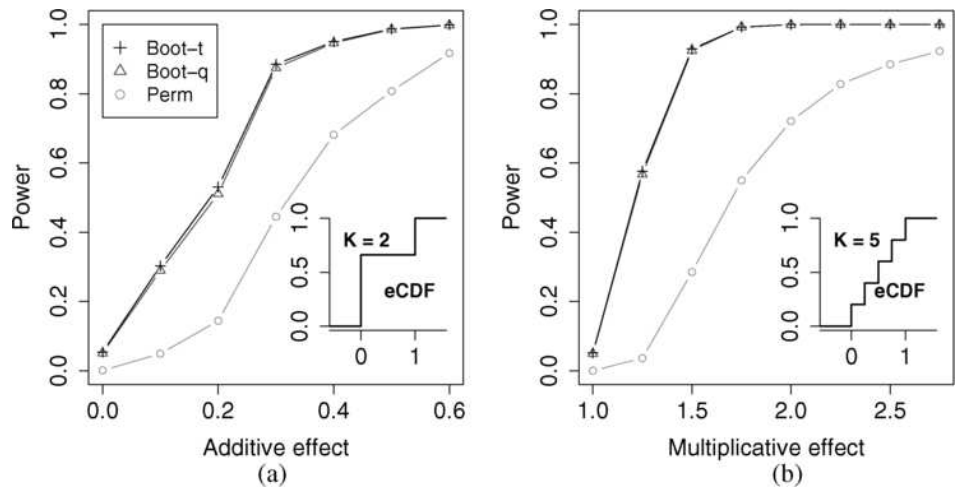

Fig. 4. *Average power of permutation- and bootstrap-based gene category tests as one departs from Class 3 nulls. Results based on randomized microarray data and real GO categories, and applying* (a) *an additive constant to $K=2$ classes of genes with $1/3$ differentially expressed at $d=1$ (as shown by the CDF in the inset graphs), and* (b) *a multiplicative constant to $K=5$ classes of genes with $\{d_k\}$ equally spaced between 0 and 1. Both scenarios exhibit more power to detect the alternative using the bootstrap tests.*

We employed the Wilcoxon rank-sum $U_W$ as a the global statistic for the permutation and bootstrap-based tests; results were obtained from 1000 permutations/resamples of the data, respectively. The $p$-values produced by

Table 2
*Number of significant GO categories for target $\alpha$ levels*

|  | **Perm** | **Boot-q** | **Boot-$t$** |
|---|---|---|---|
| $\alpha = 0.1$ | 195 | 222 | 220 |
| $\alpha = 0.05$ | 129 | 157 | 160 |
| $\alpha = 0.01$ | 56 | 72 | 85 |
| $\alpha = 0.005$ | 36 | 63 | 73 |
| $\alpha = 0.001$ | 12 | 40 | 48 |
| $\alpha = 3.7e\text{--}5^*$ | NA | NA | 28 |

*Bonferroni cutoff.

the bootstrap quantile and $t$-intervals were in good agreement across the set of categories (rank correlation $> 0.999$), reflecting the fact that the distributions of resampled global statistics were nearly normally distributed. The permutation test also showed good agreement with the bootstrap (rank correlation of 0.977 with bootstrap results), but a distinct difference was observed in the number of categories achieving various levels of significance (Table 2). The improved power of the bootstrap methods is evidenced by the increased number of significant categories, with 48 declared significant via bootstrapping at $\alpha = 0.001$, but only 12 declared significant via permutation. The minimal possible $p$-value of the permutation and bootstrap-quantile tests are limited by the 1000 resamples that were taken of the data. The bootstrap $t$-interval does not have this restriction, and 28 categories were observed to pass the conservative Bonferroni threshold for $\alpha = 0.05$. Because of the iterative procedure for estimates from the Cox-proportional hazard model, taking additional resamples of the dataset was computationally infeasible, and would be prohibitive when trying to control the FWER across such a large number of categories.

**8. Discussion.** We have used the terminology of local and global statistics as presented in SAFE [Barry, Nobel and Wright (2005)] to describe existing methods for testing differential expression within a gene category. By classifying methods according to their assumed null hypotheses, we illustrate a number of shortcomings of these methods. We propose a novel bootstrap-based approach that uniquely allows both for genes within a category and its complement to be correlated, and that maintains proper error control under a more biologically sensible null hypothesis than has been implicitly used by other methods.

As a last but very important advantage to the bootstrap-based procedure, we note that by resampling with replacement, the bootstrap can incorporate covariate information in a sensible manner. In permutation testing, by inducing a null that breaks the association between the response and expression, the covariate information can no longer be linked to both. Thus, a researcher is forced to choose the part of the data to remain linked to the covariate. By resampling the data jointly, the bootstrap allows the relationship between all three variable types to be maintained. The proper consideration of covariates is just one area of potential improvement, as gene category testing moves toward greater statistical maturity.

## APPENDIX A: PROOF TO THEOREM 1

The following elementary lemma is useful in evaluating the expectation of $U_W$.

LEMMA A.1. *Let $T_1$ and $T_2$ be distributed as $G_1$ and $G_2$ and assume that $P(T_1 = T_2) = 0$. Define $\mu(G_1, G_2) \equiv E[I\{T_1 > T_2\}]$, then $\mu(G_1, G_2) = 1 - \mu(G_2, G_1)$ and $\mu(G_1, G_2) = 1/2$ when $G_1 = G_2$.*

The expectation of $U_W$ is calculated by decomposing the $m_C \cdot m_{\bar{C}}$ pairwise comparison of $T$s into the $K^2$ different terms involving $\mu(G_k, G_{k'})$:

$$E[U_W] = E\left[\sum_{i \in C} \text{Rank}(T_i)\right] = E\left[\frac{m_C \cdot (m_C + 1)}{2} + \sum_{i \in C} \sum_{h \notin C} I\{T_i > T_h\}\right]$$

$$= \frac{m_C \cdot (m_C + 1)}{2} + \sum_{k=1}^{K} \sum_{k'=1}^{K} \sum_{\substack{i \in C \\ F_i = G_k}} \sum_{\substack{h \notin C \\ F_h = G_{k'}}} \mu(G_k, G_{k'})$$

$$= \frac{m_C \cdot (m_C + 1)}{2} + \sum_{k=1}^{K} \sum_{k'=1}^{K} m_C \cdot \beta_k \cdot m_{\bar{C}} \cdot \beta_{k'} \cdot \mu(G_k, G_{k'})$$

$$= \frac{m_C \cdot (m_C + 1)}{2}$$
$$+ m_C \cdot m_{\bar{C}} \left[\sum_{k=1}^{K} \frac{\beta_k^2}{2} + \sum_{k' < k} \beta_k \cdot \beta_{k'} [\mu(G_k, G_{k'}) + \mu(G_{k'}, G_k)]\right]$$

$$= \frac{m_C \cdot (m_C + 1)}{2} + m_C \cdot m_{\bar{C}} \left[\sum_{k=1}^{K} \frac{\beta_k^2}{2} + \sum_{k' < k} \beta_k \cdot \beta_{k'}\right]$$

$$= \frac{m_C \cdot (m_C + 1)}{2} + \frac{m_C \cdot m_{\bar{C}}}{2} \left[\sum_{k=1}^{K} \beta_k\right]^2$$

$$= \frac{m_C \cdot (m_C + 1)}{2} + \frac{m_C \cdot m_{\bar{C}}}{2} = \frac{m_C \cdot (m + 1)}{2},$$

such that $E[U_W]$ is invariant to the number of strata $K$, their proportionate sizes $\{\beta_1, \ldots, \beta_K\}$, and local statistic distributions $\{G_1, \ldots G_K\}$.

## APPENDIX B: PROOF TO THEOREM 2

The following lemma regarding the bivariate normal distribution is useful for establishing an inequality for $\text{Var}[U_W]$.

LEMMA B.2. *For the bivariate normal distribution, the following is true for the function $f(x, y) = \Phi_2(x, y; \rho) - \Phi(x) \cdot \Phi(y)$:*

1. *$f(0, 0)$ is a global maximum when $\rho > 0$,*
2. *$f(0, 0)$ is a global minimum when $\rho < 0$,*

3. $f(x, y) = 0$ when $\rho = 0$.

PROOF. The first derivatives of $f(x, y)$ are

$$\frac{\partial f}{\partial x}(x, y) = \frac{\partial}{\partial x}(\Phi_2(x, y; \rho) - \Phi(x) \cdot \Phi(y)) \propto \Phi\left(\frac{y - \rho x}{\sqrt{1 - \rho^2}}\right) - \Phi(y)$$

and $\frac{\partial f}{\partial y}$ has an analogous form due to symmetry. Since $\Phi$ is a strictly increasing function, setting the derivatives equal to zero leads to the following equations:

$$y - \rho x = \sqrt{1 - \rho^2} \cdot y,$$

$$x - \rho y = \sqrt{1 - \rho^2} \cdot x,$$

for which $\{x = 0, y = 0\}$ is the only solution when $\rho \neq 0$. Since $(0, 0)$ is the only stationary point, a second derivative test can be used to determine whether it is a global minimum or maximum [Thomas and Finney (1992)]. The second derivatives are solved to be

$$\frac{\partial^2 f}{\partial x^2}(x, y) = \phi'(x)\left[\Phi\left(\frac{y - \rho x}{\sqrt{1 - \rho^2}}\right) - \Phi(y)\right]$$

$$+ \phi(x) \cdot \phi\left(\frac{y - \rho x}{\sqrt{1 - \rho^2}}\right) \cdot \frac{-\rho}{\sqrt{1 - \rho^2}},$$

$$\frac{\partial^2 f}{\partial y^2}(x, y) = \phi'(y)\left[\Phi\left(\frac{x - \rho y}{\sqrt{1 - \rho^2}}\right) - \Phi(x)\right]$$

$$+ \phi(y) \cdot \phi\left(\frac{x - \rho y}{\sqrt{1 - \rho^2}}\right) \cdot \frac{-\rho}{\sqrt{1 - \rho^2}},$$

$$\frac{\partial^2 f}{\partial x \, \partial y}(x, y) = \phi(x) \cdot \left[\phi\left(\frac{y - \rho x}{\sqrt{1 - \rho^2}}\right) \cdot \frac{1}{\sqrt{1 - \rho^2}} - \phi(y)\right]$$

$$= \frac{\partial^2 f}{\partial y \, \partial x}(x, y).$$

At the point $\{x = 0, y = 0\}$ the derivatives are equal to

(B.1)
$$\frac{\partial^2 f}{\partial y^2}(0, 0) = \frac{\partial^2 f}{\partial x^2}(0, 0) = \phi(0)^2 \cdot \frac{-\rho}{\sqrt{1 - \rho^2}},$$

$$\frac{\partial^2 f}{\partial x \, \partial y}(0, 0) = \frac{\partial^2 f}{\partial y \, \partial x}(0, 0) = \phi(0) \cdot \left[\phi(0) \cdot \frac{1}{\sqrt{1 - \rho^2}} - \phi(0)\right]$$

and the discriminant takes the form

$$D(0, 0) = \frac{\partial^2 f}{\partial x^2}(0, 0) \cdot \frac{\partial^2 f}{\partial y^2}(0, 0) - \frac{\partial^2 f}{\partial x \, \partial y}(0, 0)^2$$

$$\begin{aligned}
&= \left(\phi(0)^2 \cdot \frac{-\rho}{\sqrt{1-\rho^2}}\right)^2 - \left(\phi(0) \cdot \left[\phi(0) \cdot \frac{1}{\sqrt{1-\rho^2}} - \phi(0)\right]\right)^2 \\
&= \phi(0)^4 \left(\frac{\rho^2}{1-\rho^2} - \frac{(1-\sqrt{1-\rho^2})^2}{1-\rho^2}\right) \\
&= \phi(0)^4 \cdot 2 \cdot \frac{\sqrt{1-\rho^2} - (1-\rho^2)}{1-\rho^2}.
\end{aligned}$$
(B.2)

Since $\sqrt{1-\rho^2} > (1-\rho^2)$ for all nonzero $\rho \in (-1,1)$, (B.2) is strictly positive, proving that either a minimum or a maximum must exist. From the second derivatives in (B.1), one can show that $f(0,0)$ is a minimum when $\rho < 0$ and a maximum when $\rho > 0$. Last, $f(x,y)$ is exactly 0 when $\rho = 0$ by independence. $\square$

The variance of $U_W$ can be decomposed using the Mann–Whitney form of the statistic:

$$\begin{aligned}
\mathrm{Var}[U_W] &= \mathrm{Var}\left[\sum_{i \in C} \mathrm{Rank}(T_i)\right] \\
&= \mathrm{Var}\left[\frac{m_C \cdot (m_C + 1)}{2} + \sum_{i \in C}\sum_{h \in \bar{C}} I\{T_i > T_h\}\right] \\
&= \sum_{i \in C}\sum_{i' \in C}\sum_{h \notin C}\sum_{h' \notin C} \mathrm{Cov}[I\{T_i > T_h\}, I\{T_{i'} > T_{h'}\}],
\end{aligned}$$
(B.3)

where

$$\begin{aligned}
&\mathrm{Cov}[I\{T_i > T_h\}, I\{T_{i'} > T_{h'}\}] \\
&= E[I\{T_i > T_h\} \cdot I\{T_{i'} > T_{h'}\}] - E[I\{T_i > T_h\}] \cdot E[I\{T_{i'} > T_{h'}\}] \\
&= P(\{T_h - T_i < 0\} \cap \{T_{h'} - T_{i'} < 0\}) - P(T_h - T_i < 0) \\
&\quad \times P(T_{h'} - T_{i'} < 0).
\end{aligned}$$

Under (6.1) and the Gaussian assumption, the paired differences in local statistics follow noncentral bivariate normal distributions with marginal means $\delta_h - \delta_i$ and $\delta_{h'} - \delta_{i'}$. Each covariance term can be written as

$$\Phi_2(\delta_h - \delta_i, \delta_{h'} - \delta_{i'}; \rho) - \Phi(\delta_h - \delta_i) \cdot \Phi(\delta_{h'} - \delta_{i'}),$$
(B.4)

where $\Phi$ and $\Phi_2$ represent the CDFs of a univariate and bivariate normal distributions with unit variance, and $\rho$ is defined as

$$\rho = \frac{\rho^T_{i,i'} + \rho^T_{h,h'} - \rho^T_{i',h} - \rho^T_{i,h'}}{\sqrt{(2 - 2\rho^T_{i,h}) \cdot (2 - 2\rho^T_{i',h'})}}.$$
(B.5)

We consider in turn the several forms of $\rho$ that arise in (B.4).

When $i = i'$ and $h = h'$, $\rho$ is proportional to $2 - 2 \cdot \rho_{i,h}^T$, which is a positive quantity except when the genes are perfectly correlated which is ruled out by the definition of a correlation dominant category. From Lemma B.2, (B.3) is maximized when $\delta_i = \delta_h$. Since this is true for all $\{i,h\}$ pairs of category and complement genes, a global maximum of the summed covariances will occur when all local statistics have the same mean.

When $i = i'$ and $h \neq h'$, $\rho$ is proportional to $1 + \rho_{i,i'}^T - \rho_{i',h}^T - \rho_{i,h'}^T$ and will be greater than 0 for a correlation dominant category such that a maximum occurs when $\delta_i = \delta_h = \delta_{h'}$. An analogous argument holds for when $i \neq i'$ and $h = h'$.

For $i \neq i'$ and $h \neq h'$, either $\rho$ will be positive if $(\rho_{i,i'}^T + \rho_{h,h'}^T) > (\rho_{i',h}^T + \rho_{i,h'}^T)$ so that (B.4) is maximized when $\delta_h = \delta_i$ and $\delta_{h'} = \delta_{i'}$, or $\rho$ will be exactly 0 if $(\rho_{i,i'}^T + \rho_{h,h'}^T) = (\rho_{i',h}^T + \rho_{i,h'}^T)$ and (B.4) will be constant. This inequality of summed correlations is again guaranteed for correlation dominant categories.

This proves a global maximum for $\text{Var}[U_W]$ is achieved at $\delta_1 = \delta_2 = \cdots = \delta_m = d$ since only in this case will every covariance term in (B.3) be either maximized, or a constant.

## SUPPLEMENTARY MATERIAL

**Supplement A: Measures of differential expression in gene category testing (Figure)** (doi: 10.1214/07-AOAS146SUPPA; .pdf). In gene category testing, global statistics typically fall into two groups: "categorical" statistics that rely on a list of significant genes to be identified, and "continuous" statistics that incorporate real-valued measures of gene-specific differential expression. The following figure illustrates the two data types using the notation framework described in the article.

**Supplement B: Variance of the Wilcoxon rank sum statistic under correlation (Theorem)** (doi: 10.1214/07-AOAS146SUPPB; .pdf). The variance of the Wilcoxon rank sum global statistic in equation (4.5) is derived in the following theorem under the assumption of dependent and identically distributed Gaussian local statistics. The proof is presented using the notation framework described in the article, and is analogous to the classic work of Gastwirth and Rubin.

## REFERENCES

Allison, D. B., Cui, X. Q., Page, G. P. et al. (2006). Microarray data analysis: From disarray to consolidation and consensus. *Nature Reviews Genetics* **7** 55–65.

Ashburner, M., Ball, C. A., Blake, J. A., Botstein, D., Butler, H., Cherry, J. M., Davis, A. P., Dolinski, K., Dwight, S. S., Eppig, J. T., Harris, M. A., Hill, D. P., Issel-Tarver, L., Kasarskis, A., Lewis, S., Matese, J. C., Richardson, J. E., Ringwald, M., Rubin, G. M. and Sherlock, G. (2000). Gene Ontology:

Tool for the unification of biology. The Gene Ontology Consortium. *Nat. Genet.* **25** 25–29.

BARRY, W. T., NOBEL, A. B. and WRIGHT, F. A. (2005). Significance analysis of functional categories in gene expression studies: A structured permutation approach. *Bioinformatics* **21** 1943–1949.

BARRY, W. T., NOBEL, A. B. and WRIGHT, F. A. (2008). Supplement to "A statistical framework for testing functional categories in microarray data." DOI: 10.1214/07-AOAS146SUPPA, DOI: 10.1214/07-AOAS146SUPPB.

BEISSBARTH, T. and SPEED, T. P. (2004). GOstat: Find statistically overrepresented Gene Ontologies within a group of genes. *Bioinformatics* **20** 1464–1465.

BEN-SHAUL, Y., BERGMAN, H. and SOREQ, H. (2005). Identifying subtle interrelated changes in functional gene categories using continuous measures of gene expression. *Bioinformatics* **21** 1129–1137.

BHATTACHARJEE, A., RICHARDS, W. G., STAUNTON, J., LI, C., MONTI, S., VASA, P., LADD, C., BEHESHTI, J., BUENO, R., GILLETTE, M., LODA, M., WEBER, G., MARK, E. J., LANDER, E. S., WONG, W., JOHNSON, B. E., GOLUB, T. R., SUGARBAKER, D. J. and MEYERSON, M. (2001). Classification of human lung carcinomas by mRNA expression profiling reveals distinct adenocarcinoma subclasses. *Proc. Natl. Acad. Sci. USA* **98** 13790–13795.

BOORSMA, A., FOAT, B. C., VIS, D., KLIS, F. and BUSSEMAKER, H. J. (2005). T-profiler: scoring the activity of predefined groups of genes using gene expression data. *Nucleic Acids Research* **33** W592–W595.

CASELLA, G. and BERGER, R. L. (2002). *Statistical Inference*, 2nd ed. Duxbury, Australia. MR1051420

CHANG, H. Y., NUYTEN, D. S. A., SNEDDON, J. B., HASTIE, T., TIBSHIRANI, R., SORLIE, T., DAI, H. Y., HE, Y. D. D., VEER, L. J. V., BARTELINK, H., DE RIJN, M. V., BROWN, P. O. and DE VIJVER, M. J. V. (2005). Robustness, scalability, and integration of a wound-response gene expression signature in predicting breast cancer survival. *Proc. Natl. Acad. Sci. USA* **102** 3738–3743.

DAMIAN, D. and GORFINE, M. (2004). Statistical concerns about the GSEA procedure. *Nature Genetics* **36** 663.

DODD, L. E., SENGUPTA, S., CHEN, I. H., DEN BOON, J. A., CHENG, Y. J., WESTRA, W., NEWTON, M. A., MITTL, B. F., MCSHANE, L., CHEN, C. J., AHLQUIST, P. and HILDESHEIM, A. (2006). Genes involved in DNA repair and nitrosamine metabolism and those located on chromosome 14q32 are dysregulated in nasopharyngeal carcinoma. *Cancer Epidemiology Biomarkers and Prevention* **15** 2216–2225.

DRAGHICI, S., KHATRI, P., MARTINS, R. P., OSTERMEIER, G. C. and KRAWETZ, S. A. (2003). Global functional profiling of gene expression. *Genomics* **81** 98–104.

DUDOIT, S., KELES, S. and VAN DER LAAN, M. J. (2007). *Multiple Tests of Association with Biological Annotation Metadata.* Springer, New York.

DUDOIT, S., YANG, Y. H., CALLOW, M. J. and SPEED, T. P. (2002). Statistical methods for identifying differentially expressed genes in replicated cDNA microarray experiments. *Statist. Sinica* **12** 111–139. MR1894191

EFRON, B. (1979). Bootstrap methods: Another look at the jackknife. *Ann. Statist.* **7** 1–26. MR0515681

EFRON, B. (1987). Better bootstrap confidence intervals. *J. Amer. Statist. Assoc.* **82** 171–185. MR0883345

EFRON, B. and TIBSHIRANI, R. J. (1998). *An Introduction to the Bootstrap*, 2nd ed. Chapman and Hall/CRC, New York. MR1270903

Efron, B. and Tibshirani, R. (2007). On testing the significance of sets of genes. *Ann. Applied Statist.* **1** 107–129.

Galitski, T., Saldanha, A. J., Styles, C. A., Lander, E. S. and Fink, G. R. (1999). Ploidy regulation of gene expression. *Science* **285** 251–254.

Gastwirth, J. L. and Rubin, H. (1971). Effect of dependence on the level of some one-sample tests. *J. Amer. Statist. Assoc.* **66** 816–820. MR0314192

Goeman, J. J. and Buhlmann, P. (2007). Analyzing gene expression data in terms of gene sets: Methodological issues. *Bioinformatics* **23** 980–987.

Hall, P. and Wilson, S. R. (1991). Two guidelines for bootstrap hypothesis testing. *Biometrics* **47** 757–762. MR1132543

Kim, S.-Y. and Volsky, D. J. (2005). Parametric analysis of gene set enrichment. *BMC Bioinformatics* **6** 144.

Lee, H. K., Hsu, A. K., Sajdak, J., Qin, J. and Pavlidis, P. (2004). Coexpression analysis of human genes across many microarray data sets. *Genome Research* **14** 1085–1094.

Mootha, V. K., Lindgren, C. M., Eriksson, K. F., Subramanian, A., Sihag, S., Lehar, J., Puigserver, P., Carlsson, E., Ridderstrale, M., Laurila, E., Houstis, N., Daly, M. J., Patterson, N., Mesirov, J. P., Golub, T. R., Tamayo, P., Spiegelman, B., Lander, E. S., Hirschhorn, J. N., Altshuler, D. and Groop, L. C. (2003). PGC-1alpha-responsive genes involved in oxidative phosphorylation are coordinately downregulated in human diabetes. *Nat. Genet.* **34** 267–273.

Newton, M. A., Noueiry, A., Sarkar, D. and Ahlquist, P. (2004). Detecting differential gene expression with a semiparametric hierarchical mixture method. *Biostatistics* **5** 155–176.

Pavlidis, P., Qin, J., Arango, V., Mann, J. J. and Sibille, E. (2004). Using the gene ontology for microarray data mining: A comparison of methods and application to age effects in human prefrontal cortex. *Neurochemical Research* **29** 1213–1222.

Pearson, K. (1911). on the probability that two independent distributions of frequency are really samples from the same population. *Biometrika* **8** 250–254.

Subramanian, A., Tamayo, P., Mootha, V. K., Mukherjee, S., Ebert, B. L., Gillette, M. A., Paulovich, A., Pomeroy, S. L., Golub, T. R., Lander, E. S. and Mesirov, J. P. (2005). Gene set enrichment analysis: A knowledge-based approach for interpreting genome-wide expression profiles. *Proc. Natl. Acad. Sci. USA* **102** 15545–15550.

Thomas, G. B. J. and Finney, R. L. (1992). *Maxima, Minima, and Saddle Points*, 8th ed. Addison-Wesley, Reading, MA.

Tusher, V. G., Tibshirani, R. and Chu, G. (2001). Significance analysis of microarrays applied to the ionizing radiation response. *Proc. Natl. Acad. Sci. USA* **98** 5116–5121.

Virtaneva, K. I., Wright, F. A., Tanner, S. M., Yuan, B., Lemon, W. J., Caligiuri, M. A., Bloomfield, C. D., de la Chapelle, A. and Krahe, R. (2001). Expression profiling reveals fundamental biological differences in acute myeloid leukemia with isolated trisomy 8 and normal cytogenetics. *Proc. Natl. Acad. Sci. USA* **98** 1124–1129.

Zhong, S., Storch, K. F., Lipan, O., Kao, M. C., Weitz, C. J. and Wong, W. H. (2004). GoSurfer: A graphical interactive tool for comparative analysis of large gene sets in gene ontology space. *Appl. Bioinformatics* **3** 261–264.

W. T. BARRY
DEPARTMENT OF BIOSTATISTICS
  AND BIOINFORMATICS
DUKE UNIVERSITY MEDICAL CENTER
DURHAM, NORTH CAROLINA 27710
USA
E-MAIL: bill.barry@duke.edu

A. B. NOBEL
DEPARTMENT OF STATISTICS
  AND OPERATIONS RESEARCH
UNIVERSITY OF NORTH CAROLINA AT CHAPEL HILL
CHAPEL HILL, NORTH CAROLINA 27599-3260
USA
E-MAIL: nobel@email.unc.edu

F. A. WRIGHT
DEPARTMENT OF BIOSTATISTICS
UNIVERSITY OF NORTH CAROLINA AT CHAPEL HILL
CHAPEL HILL, NORTH CAROLINA 25599-7420
USA
E-MAIL: fwright@bios.unc.edu